\documentclass[final,3p,times,twocolumn]{elsarticle}

\usepackage{graphics}

\usepackage{amssymb}
\usepackage{amsmath}
\usepackage{physics}
\usepackage{textcomp}
\usepackage{url}
\usepackage{tabularx}
\usepackage{lineno}

\journal{JQSRT}

\newcommand{\ee}[1]{$\cdot$10$^{#1}$}
\newcommand{\degree}{\ensuremath{^\circ}}

\begin{document}

\begin{frontmatter}



\title{Errors induced by the neglect of polarization in radiance
  calculations for three-dimensional cloudy atmospheres}
\author{Claudia Emde}
\ead{claudia.emde@lmu.de}
\author{Bernhard Mayer}

\address{Meteorological Institute, Ludwig-Maximilians-University (LMU),
  Theresienstr. 37, Munich, Germany}

\begin{abstract}

Remote sensing instruments observe radiation being scattered
and absorbed by molecules, aerosol particles, cloud droplets and ice
crystals. In order to interpret and accurately model such observations, the vector 
radiative transfer equation needs to be solved,
because scattering polarizes the
initially unpolarized incoming solar radiation.  
A widely used approximation in radiative transfer theory is the
neglect of polarization which allows to greatly simplify the radiative
transfer equation.

It is well known that the error caused by multiple Rayleigh scattering
can be larger than 10\%, depending on wavelength and sun-observer
geometry \citep{mishchenko1994}. For homogeneous plane-parallel layers of liquid cloud droplets the error is
comparatively small (below 1\%) 
\citep{hansen1971}. Therefore, in radiative transfer modelling for 
cloud remote sensing polarization is mostly neglected.
However, in reality clouds are not plane-parallel
layers of water droplets but complex three-dimensional (3D)
structures and observations of clouds usually include
pixels consisting of clear and cloudy parts. In
this study we revisit the question of the magnitude of
error due to the neglect of polarization in radiative transfer theory for a realistic
3D cloudy atmosphere.

We apply the Monte Carlo radiative transfer model MYSTIC
with and without neglecting polarization and
compare the results. At a phase angle of 90\degree\ and 400\,nm wavelength
we find the maximum overestimation error of about
8\% for complete clear-sky conditions. The error is reduced to about 6\% in
clear-sky regions surrounded by clouds due to scattering from
clouds into the clear regions. 
Within the clouds the error is up to 4\% with the highest values in
cloud shadows. 
In backscattering direction the radiance is underestimated by
about 5\% in clear regions between clouds. 
For other sun-observer geometries, the error ranges
between the two extremes. The error decreases with wavelength and in
the absorption bands.


\end{abstract}

\begin{keyword}
3D radiative transfer \sep polarization \sep
scalar approximation \sep cloud scattering \sep Rayleigh scattering 

\end{keyword}

\end{frontmatter}


\section{Introduction}
\label{sec:intro}

The vector radiative transfer
equation is an integro-differential equation for the so-called Stokes
vector \citep{chandrasekhar50}, which can be solved numerically using
a variety of approaches (e.g. , the doubling-and-adding method
\citep{hansen1974,dehaan1987}, 
the spherical harmonics discrete ordinate method \citep{doicu2013}
or the Monte Carlo method \citep{collins1972, emde2010, cornet2010}).
A commonly used approximation to the
rigorous vector radiative transfer equation is the scalar radiative
transfer equation, which simplifies the 
numerical solution and allows much faster calculations of radiances. 
Widely used radiative transfer codes as for
example DISORT (discrete ordinate method
\citep{stamnes2000,buras2011b})
use the scalar approximation. 

In his pioneering book, \citet{chandrasekhar50} already pointed out 
that accurate radiative transfer calculations for Rayleigh
scattering need to consider that radiation is polarized. 
Several studies followed and consistently found errors above
10\% for light reflected by pure Rayleigh scattering layers
(e.g. \citet{adams1970}, \citet{mishchenko1994} and references
therein, \citet{Kotchenova2006}). 

The most detailed and comprehensive work by 
\citet{mishchenko1994} has shown, that for clear-sky molecular atmospheres
the error due to the neglect of polarization is larger than 10\%
when the Rayleigh scattering optical thickness is about
1 for phase angles close to 90\degree\ and in backscattering
directions. The phase angle is the defined as the angle between
incident solar radiation and viewing direction.
\citet{mishchenko1994} have 
performed extensive simulations for various optical thicknesses,
single scattering albedos, surface albedos, Rayleigh depolarization factors and
sun-observer geometries to provide the information, under which
conditions it is important to use the full vector radiative
transfer equation.

\citet{hansen1971} examined the error of the scalar approximation for
particles as large as the wavelength or larger by simulating radiances
for a plane-parallel layer including cloud particles. He found the
largest error for a cloud optical thickness of about 1. At wavelengths
of 1.2\,$\mu$m and 2.25\,$\mu$m it was typically about 0.1\%, and at
3.4\,$\mu$m the error was commonly about 1\%.  \citet{hansen1971}
explains that radiances calculated using the scalar approximation are
much more accurate for scattering in clouds than for Rayleigh
scattering because of the greater amount of forward scattering and the
smaller polarization by single scattering. The largest error results
from ``photons''\footnote{We use the term ``photon'' to represent a
discrete amount of electromagnetic energy transported in a
specific direction. It is not related to the QED photon
\citep{mishchenko2014}.} that are scattered a few times (more than
once). For an optically thick cloud the number of photons being
scattered a few times is small compared to total number of photons
contributing to the radiance. This argument holds for spherical and
non-spherical particles, thus \citet{hansen1971} concludes that in
most cases the error of the scalar approximation should be smaller
than 1\% for reflection from a cloud of particles which are at least
as large as the wavelength.

For aerosol scattering, \citet{Kotchenova2006} find errors up to about
5\% at 670\,nm
wavelength for biomass burning aerosol with an optical thickess of
about 0.7. The assumed aerosol size distribution is dominated by small
particles with a mean radius of approximately 0.15\,$\mu$m, thus
significant errors due to the neglect of polarization can be expected
since the mean particle size is smaller than the wavelength of the
radiation. In this size range the shape of the aerosol particles also
has a significant impact on the polarization (e.g.,
\citep{waquet2007}), thus the error might also depend on the particle
shape and particle orientation.
\citet{Barlakas2016b} calculated the error at 532\,nm for
larger mineral dust particles with strong forward scattering
and found errors below 1\%. 

Since for liquid water and ice clouds the particle size is always larger than the
wavelength in optical remote sensing, polarization is commonly
neglected. Algorithms to retrieve cloud microphysical properties often
rely on lookup-tables including pre-calculated (scalar) radiances
(e.g. \citep{nakajima1990,zinner2010,zinner2016}).  An advanced tomographic
retrieval method \citep{levis2015,levis2017} which
considers the 3D structure of the clouds applies the 3D scalar model 
SHDOM by \citet{Evans1998}. So far, polarization is only considered in
cloud remote sensing algorithms for polarized observations
(e.g. POLDER \citep{deschamps1994,breon2005,stap2016b}). 

\citet{yi2014} investigated the influence of polarization on
cloud-property retrievals from Moderate Resolution Imaging
Spectroradiometer (MODIS) satellite observations and found differences
in retrieved water and ice cloud effective radius and optical thickness
differences of as much as $\pm$15\%. They attribute those differences
to the neglect of polarization in the one-dimensional forward model
calculations. 

In this study, we revisit the question of whether it is important to
consider polarization in radiative transfer simulations for cloud
remote sensing. In particular for coarse resolution satellite
observations, coarse pixels may actually be partially cloudy -- a
fraction of the pixel may be clear-sky. Additionally, the cloud itself
is often not horizontally homogeneous and significant cloud top height
inhomogeneity can result in cloud shadowing effects. Together these
effects result in observed cloudy pixels that include a
significant Rayleigh scattering contribution.
In order to calculate the error due to the neglect of polarization, 
we performed radiative transfer simulations using the Monte
Carlo model MYSTIC \citep{mayer2009,emde2010}
for a realistic three-dimensional cloud field generated by a Large
Eddy Simulation (LES) model. We run MYSTIC in full
vector mode and in scalar approximation mode, respectively, 
and by comparing the
results, we obtain the error due to the neglect of polarization. In
order to mimic satellite observations, we spatially average the results to obtain
coarser spatial resolutions including partially cloudy pixels. 

The paper is organized as follows: In Section~\ref{sec:rte} 
the vector radiative transfer equation and its scalar approximation
are given. 
Section~\ref{sec:methodology} briefly describes the radiative transfer
model and the atmospheric input parameters. Section~\ref{sec:results} shows
the error of the scalar approximation for the various setups. Finally,
in
Section~\ref{sec:conclusions} we summarize the results and discuss their
importance for cloud remote sensing. 

\section{Radiative transfer equation}
\label{sec:rte}

Multiple scattering and absorption of solar radiation in the
atmosphere is described by the matrix integro-differential equation
called the vector radiative transfer equation (VRTE). A
phenomenological derivation of the VRTE can be found in
e.g. \citet{hansen1974}. \citet{mishchenko2006} rigorously derived the VRTE for a
plane-parallel medium from Maxwell's equations. 
\citet{mishchenko2006b} has generalized the derivation to clouds
with inhomogeneities that are small with respect to the photon's
mean-free-path, which is not necessarily fulfilled in realistic
three-dimensional cloud fields. Unfortunately, 
a rigorous derivation of the three-dimensional
VRTE from Maxwell's equations for inhomogeneous media is not yet available. We
will nevertheless use the phenomenologically derived VRTE in the
following form:
\begin{align}
  \label{eq:vrte}
  (\hat{\bf n} \cdot \grad) {\bf I} (\hat{\bf n}, {\bf x})  = &
  \nonumber \\  
  -{\bf K} ({\bf x}){\bf I} (\hat{\bf n}, {\bf x})  +  &
  \frac{k_{\rm sca}}{4\pi} \int_{4\pi}\diffd \hat{\bf n}'
  {\bf L}(\sigma_2){\bf P}(\Theta, {\bf x}){\bf L}(\sigma_1) 
  {\bf I}(\hat{\bf n}', {\bf x})  
\end{align}
where $\hat{\bf n}$ is
the propagation direction of the radiation,
${\bf x}$ is the position vector in the three-dimensional medium,
${\bf K}$ is the extinction matrix, $k_{\rm sca}$ is the scattering coefficient, 
${\bf P}$ is the
scattering phase matrix, which for spherical or randomly oriented
particles depends only on the scattering angle
$\Theta=\arccos(\hat{\bf n}\cdot\hat{\bf n}')$,
and ${\bf L}$ is the Stokes rotation matrix (see e.g. \citet{mishchenko1994}),
which transforms the Stokes vector ${\bf I}$
from its reference frame to the scattering frame and vice versa. 
The reference frame of the Stokes vector is defined by the zenith
direction and the propagation direction of the radiation. The
scattering frame is defined by the incoming and outgoing directions.
The angles $\sigma_1$ and $\sigma_2$ are the angles between scattering
frame and reference frames of the Stokes vector before and after being
scattered. 
${\bf I}$ is the four component Stokes vector defined as follows:
\begin{eqnarray}
  {\bf I} =
  \begin{pmatrix}
    I \\ Q \\ U \\ V
  \end{pmatrix} 
  = \frac{1}{2}\sqrt{\frac{\epsilon}{\mu_p}}
  \begin{pmatrix}
   \langle E_\parallel E_\parallel^\ast + E_\perp E_\perp^\ast \rangle\\
   \langle E_\parallel E_\parallel^\ast - E_\perp E_\perp^\ast \rangle\\
   \langle -E_\parallel E_\perp^\ast - E_\perp E_\parallel^\ast \rangle\\
   \langle i(E_\parallel E_\perp^\ast- E_\perp E_\parallel^\ast) \rangle 
  \end{pmatrix}
  \label{eq:stokes}
\end{eqnarray}
Here, $E_\parallel$ and $E_\perp$ are the components of the electric field vector
parallel and perpendicular to
the reference plane, respectively. The brackets $\langle$ ... $\rangle$ denote that the electric field is averaged over random realizations of the phases, i.e. over a certain space-time domain.  The pre-factor on the right hand
side contains the electric permittivity $\epsilon$ and the magnetic
permeability $\mu_p$.

Often, only the scalar radiance, i.e. the first Stokes component $I$ is
required. In order to calculate this, the radiative transfer equation can be
approximated in scalar form: 
\begin{equation}
  \label{eq:srte}
  (\hat{\bf n} \cdot \grad) I (\hat{\bf n}, {\bf x})= -k_{\rm ext}({\bf
    x}) I(\hat{\bf n}, {\bf x}) +  \frac{k_{\rm sca}}{4\pi}
  \int_{4\pi}\diffd \hat{\bf n}' P(\Theta, {\bf x}) I (\hat{\bf  n}', {\bf x})  
\end{equation}
Here $k_{\rm ext}$ is the $(1,1)$-element of the extinction matrix and 
$P(\Theta, {\bf x})$ is the scattering phase function, which
corresponds to the $(1,1)$-element of the scattering phase matrix.

In this study we investigate the error $ E(\hat{\bf n}, {\bf x})$ 
induced by the scalar
approximation of the radiative transfer equation: 
\begin{equation}
  \label{eq:error}
  E(\hat{\bf n}, {\bf x}) = \frac{ I^v 
    (\hat{\bf n}, {\bf x})- I^s (\hat{\bf n}, {\bf x}) }
  {I^v(\hat{\bf n}, {\bf x})}
\end{equation}
$I^v(\hat{\bf n}, {\bf x})$ is the radiance calculated by solving the
vector radiative transfer Eq.~\ref{eq:vrte} and $I^s (\hat{\bf
  n}, {\bf x})$ is calculated using its scalar approximation
Eq.~\ref{eq:srte}.

\section{Methodology and model setup}
\label{sec:methodology}

\subsection{Radiative transfer model MYSTIC}

We apply the libRadtran \citep{emde2016,mayer2005} radiative transfer
model with the Monte Carlo solver MYSTIC \citep{mayer2009,emde2010},
which can be run in scalar mode
neglecting polarization, and in vector mode, where polarization is
fully considered. For both runs we use exactly the same input, the
only difference is, that in vector mode, the full phase matrix is used
when photons
are scattered whereas in scalar mode, only the $P_{11}$
element of the phase matrix is used. The implementation of
polarization in MYSTIC is described in detail in \citet{emde2010}. 
Since highly peaked phase matrices cause convergence problems in Monte
Carlo simulations, we apply sophisticated variance reduction
techniques \citep{buras2011}, which work in the same way 
for scalar and vector simulations.   

MYSTIC applies periodic boundary conditions in $x$- and $y$-direction,
which means that photons that leave the domain
on one side at a certain vertical position 
re-enter the domain on the opposite side at the same vertical position
without changing their propagation direction. 

MYSTIC has participated in various model intercomparison studies
\citep{cahalan2005,Kokhanovsky2010,emde2015,emde2018} and always
agreed perfectly to commonly established benchmark results.


\subsection{Atmosphere and cloud definition}

As atmospheric background we use the US standard atmosphere as defined
by \citet{anderson1986}. Molecular scattering coefficients and the Rayleigh
depolarization factor $\delta$, that accounts for the
anisotropy of the molecules, are calculated according
to \citet{Bodhaine1999}. For molecular absorption we used the
REPTRAN absorption parameterization \citep{gasteiger2014}. At 400\,nm
we obtain an integrated Rayleigh scattering optical thickness of 0.36
and a small molecular absorption optical thickness of 3.7\ee{-3}. 
The Rayleigh depolarization factor at 400\,nm is 0.03. 
Rayleigh scattering is characterized by the phase matrix
\citep{hansen1974}:
 \begin{center}
\begin{eqnarray}
  \label{eq:rayleigh_phase_matrix}
  \begin{array}{l}
   {\bf P}_{\rm R}(\Theta) =  \\[2ex]
   \Delta \left[ 
    \begin{array}{cccc}
      \frac{3}{4}(1+\cos^2\Theta) & -\frac{3}{4}\sin^2\Theta &  0 & 0 \\
      -\frac{3}{4}\sin^2\Theta & \frac{3}{4}(1+\cos^2\Theta) &  0 & 0 \\
      0 & 0  & \frac{3}{2}\cos \Theta & 0 \\
      0 & 0 & 0 & \Delta' \frac{3}{2}\cos \Theta 
    \end{array}
  \right] \nonumber \\[6ex]
\hspace{10ex}+(1-\Delta)\left[
\begin{array}{cccc}
 \ 1\ &\ 0\ &\ 0\ &\ 0\ \\
 \ 0\ &\ 0\ &\ 0\ &\ 0\ \\
 \ 0\ &\ 0\ &\ 0\ &\ 0\ \\
 \ 0\ &\ 0\ &\ 0\ &\ 1\ 
\end{array}
\ \right],
\end{array}
\end{eqnarray}
\end{center} 
where
\begin{eqnarray}
  \label{eq:depol}
  \Delta=\frac{1-\delta}{1+\delta/2}, \qquad 
  \Delta'=\frac{1-2\delta}{1-\delta},
\end{eqnarray}
and $\delta$ is the depolarization factor and $\Theta$ is the
scattering angle. The elements of the Rayleigh phase matrix are shown
in Figure~\ref{fig:phase_matrix}.

We include a shallow cumulus cloud field from large-eddy simulations
(LES) by \citet{stevens1999} in the model domain. This cloud field has
already been used for various model intercomparison studies
\citep{cahalan2005,emde2018}. The cloud field includes
100$\times$100$\times$36 grid cells with a size of
66.7$\times$66.7$\times$40\,m$^3$. For each grid cell the cloud extinction
coefficient and the effective droplet radius is given. 
\begin{figure}[htbp]
  \centering
  \includegraphics[width=0.9\hsize]{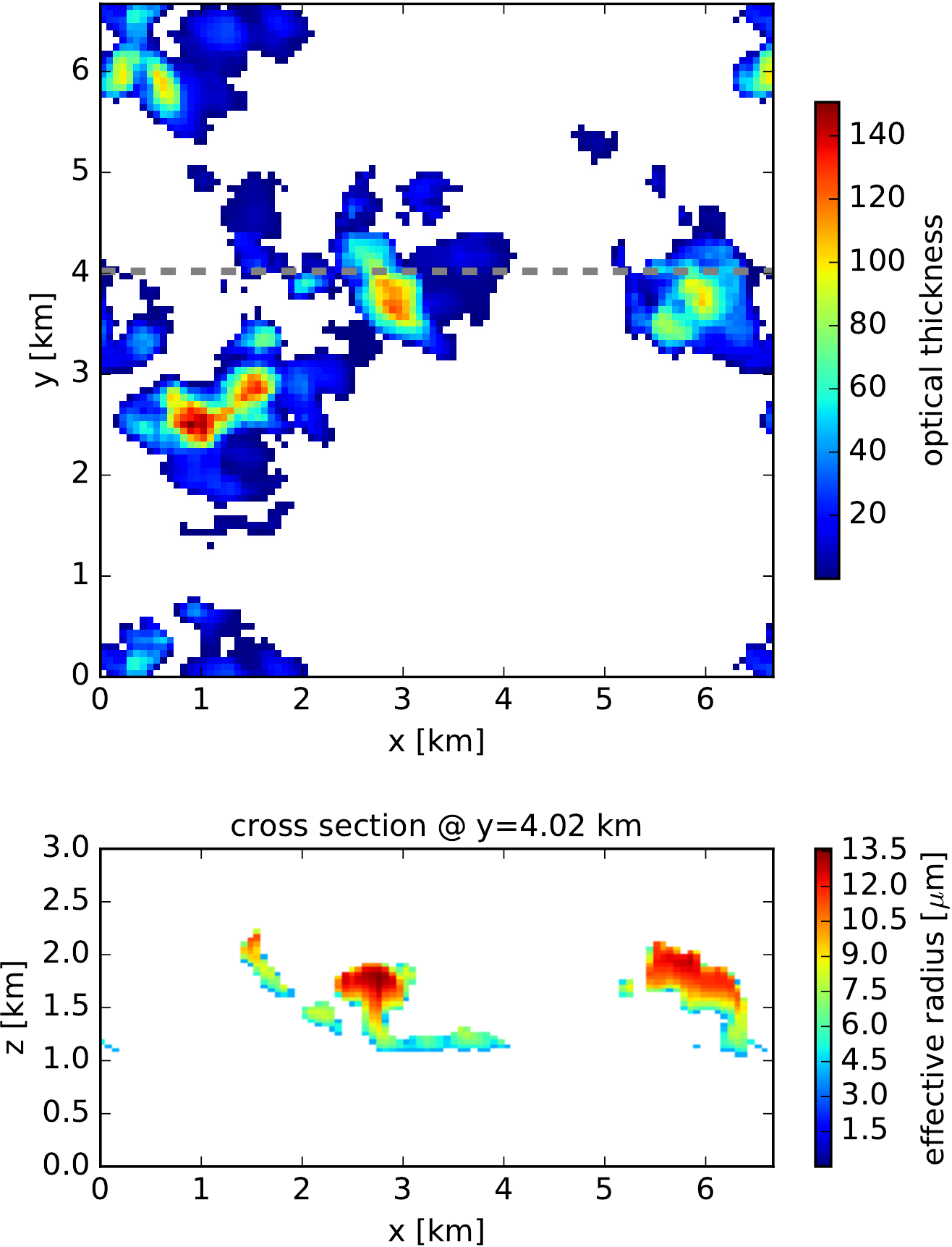}
  \caption{Definition of cumulus cloud field. The upper figure shows the
    integrated vertical optical thickness and the lower figure shows
    the effective radius for a
    vertical cross section (gray dashed line in upper figure)
    through the domain at $y$\,=\,4.02\,km.}
  \label{fig:cumulus_cloud_def}
\end{figure}
The upper panel in Figure~\ref{fig:cumulus_cloud_def}
shows the vertically integrated optical thickness of
the clouds and the lower panel shows the effective radius
for a vertical cross section at y\,=\,4.02\,km. 

The cloud optical properties have been computed using Mie
theory \citep{mie1908,wiscombe80a}. We assumed a
gamma size distribution with an effective variance of 0.1, a value
typical for liquid water clouds.  
The phase matrix $ {\bf P}_{\rm C}$ for spherical particles has only 4 independent
elements and it has the following form:
\begin{equation}
  {\bf P}_{\rm C}(\Theta) = \left[
\begin{array}{cccc}
  P_{11}(\Theta) & P_{12}(\Theta) &  0 & 0 \\
  P_{12}(\Theta) & P_{11}(\Theta) &  0 & 0 \\
  0 & 0 & P_{33}(\Theta) & P_{34}(\Theta) \\
  0 & 0 & -P_{34}(\Theta) &  P_{33}(\Theta)
  \label{eq:scat_matrix_rand}
\end{array}
\right]
\end{equation}
Figure~\ref{fig:phase_matrix} shows the scattering phase matrix at 400\,nm for
effective radii of 5\,$\mu$m, 10\,$\mu$m, and 15\,$\mu$m, 
this is roughly the range we find
in the cumulus cloud field.  
\begin{figure}[htbp]
  \centering
  \includegraphics[width=1.0\hsize]{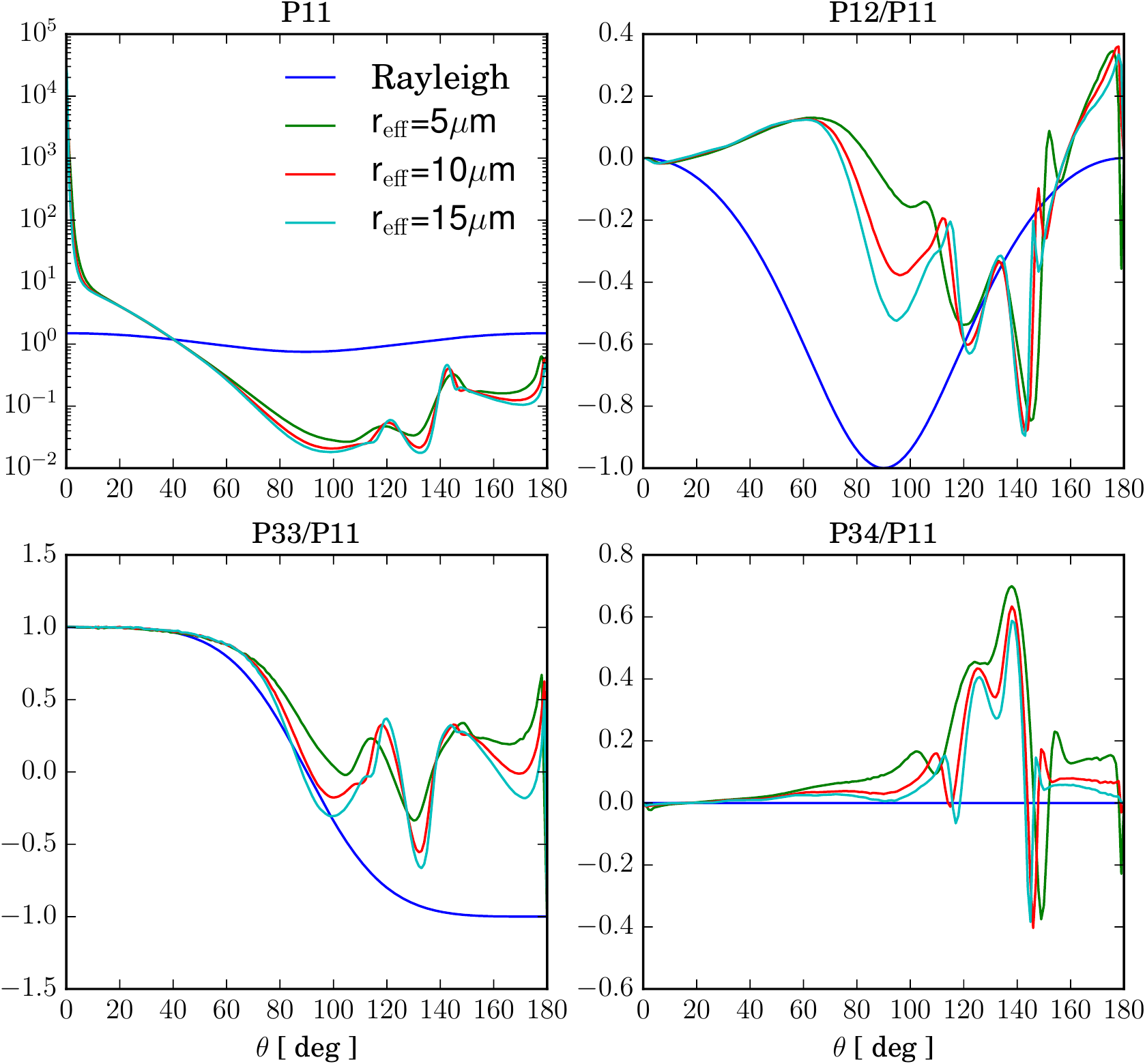}
  \caption{Phase matrix for Rayleigh scattering and liquid water cloud scattering.}
  \label{fig:phase_matrix}
\end{figure}

We assumed that the surface is purely absorbing, i.e. the surface albedo was set to zero. 

\section{Results}
\label{sec:results}

\subsection{Results for one-dimensional clear atmosphere}

\begin{figure*}
  \centering
  \includegraphics[width=1.0\hsize]{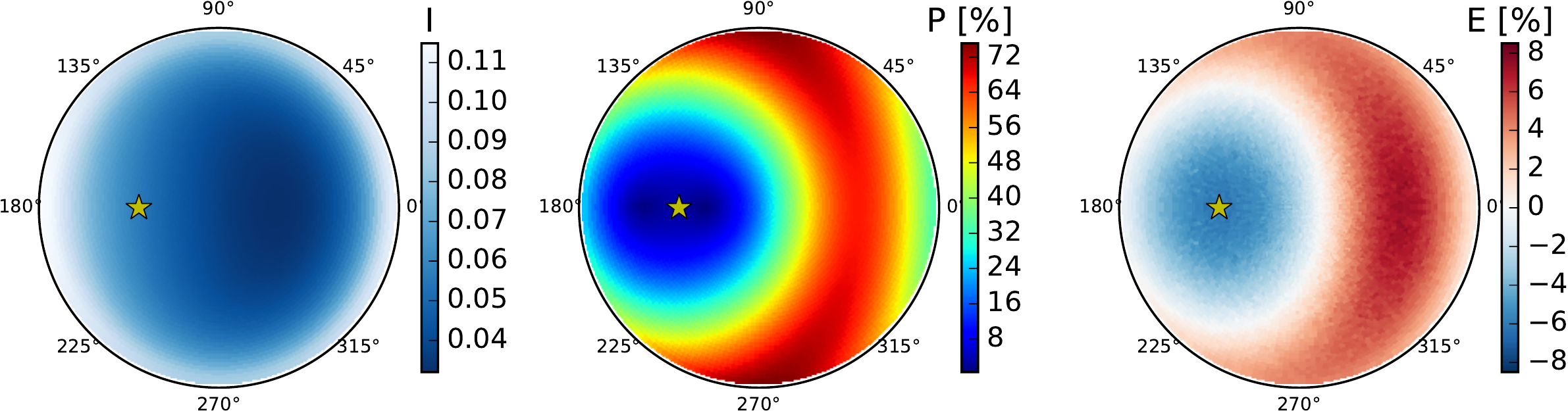}
  \caption{Normalized radiance, degree of polarization and error of scalar
    approximation at a wavelenth of 400\,nm for the US-standard atmosphere for 
    a down-looking observer at 120\,km altitude. The
    sun position is ($\theta_0, \phi_0$)=(40\degree, 0\degree) and the
    yellow star shows the viewing direction where the sun is in the back
    of the sensor ($\theta, \phi$)=(140\degree, 180\degree). 
    The maximum overestimation is obtained for phase
    angles around 90\degree, and the maximum underestimation for
    backscattering directions close to a phase angle of 0\degree.}
  \label{fig:error_polarplot}
\end{figure*}
In order to provide an overview of the error due to neglecting
polarization depending on viewing direction we calculate the radiance
field at 120\,km altitude (top of US-standard atmosphere)
for a solar zenith angle $\theta_0$ of
40\degree\ and a solar azimuth angle $\phi_0$ of 0\degree.
The radiance in this and all other simulations has been normalized to
the extraterrestrial irradiance.  
The left plot in Figure~\ref{fig:error_polarplot} shows the 
radiance
for all viewing directions as a polar
plot for the US-standard atmosphere at a wavelength of 400\,nm.
We see the typical Rayleigh scattering pattern with
increasing radiances towards the horizon and smallest radiance values
for directions, where the angle between incident
solar radiation and viewing direction, the so-called phase angle, is
about 90\degree. 
The middle plot shows the
corresponding degree of polarization, which is smallest around the
backscattering direction and largest at a phase angle of
90\degree. The right plot shows the error due to the scalar
approximation. We find an overestimation of up to about 8\% at
a phase angle of about 90\degree\ and an underestimation of up to
about 7\% in the backscattering region. These results are consistent
with \citet{mishchenko1994}, who explains the under- and
overestimation by second-order and further low-order scattering
processes. 
We also calculated the error for
a sensor at the surface. The pattern of the results (not shown)
is very similar to that for the sensor at the top of the atmosphere. 
For phase angles about
90\degree\ the maximum overestimation of about 8\% is obtained. In
forward scattering directions, when the observer ``looks'' in a
direction close to the sun, the
radiance is underestimated by about 7\%. 

\begin{figure*}
  \centering
  \includegraphics[width=1.0\hsize]{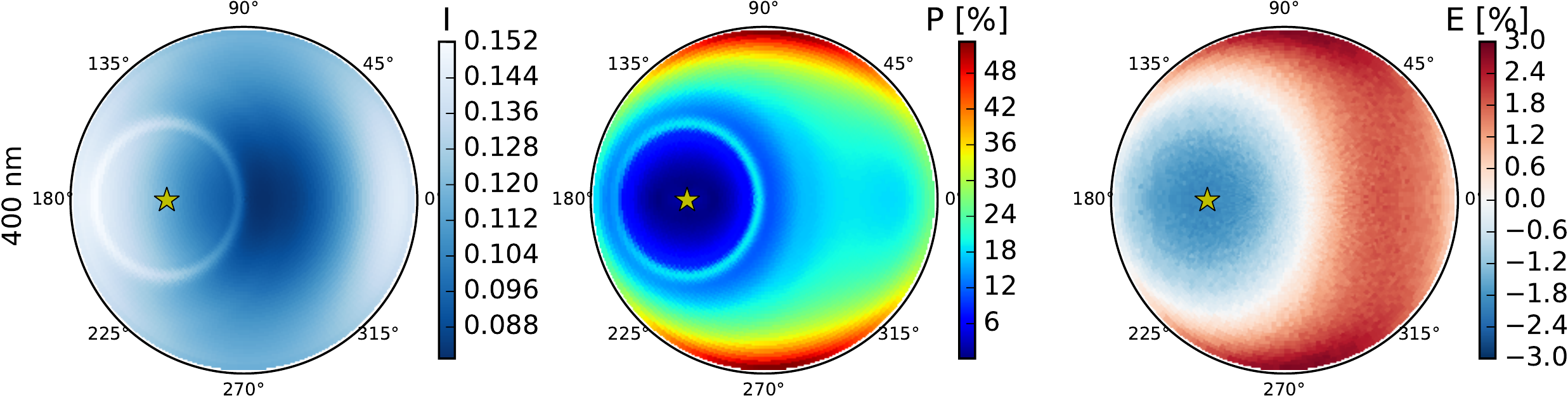}\\
  \includegraphics[width=1.0\hsize]{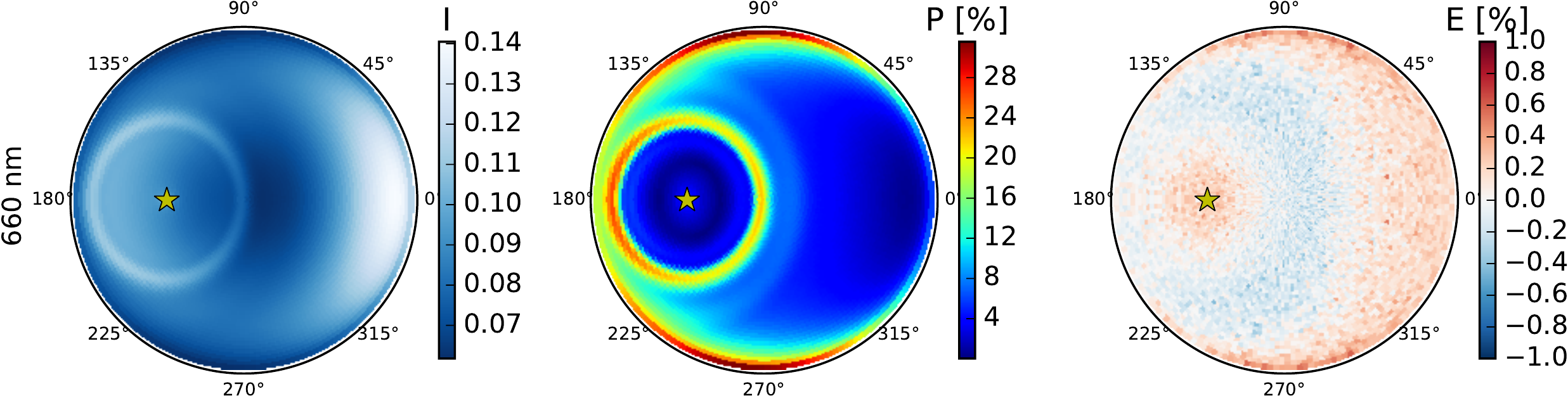}\\
  \includegraphics[width=1.0\hsize]{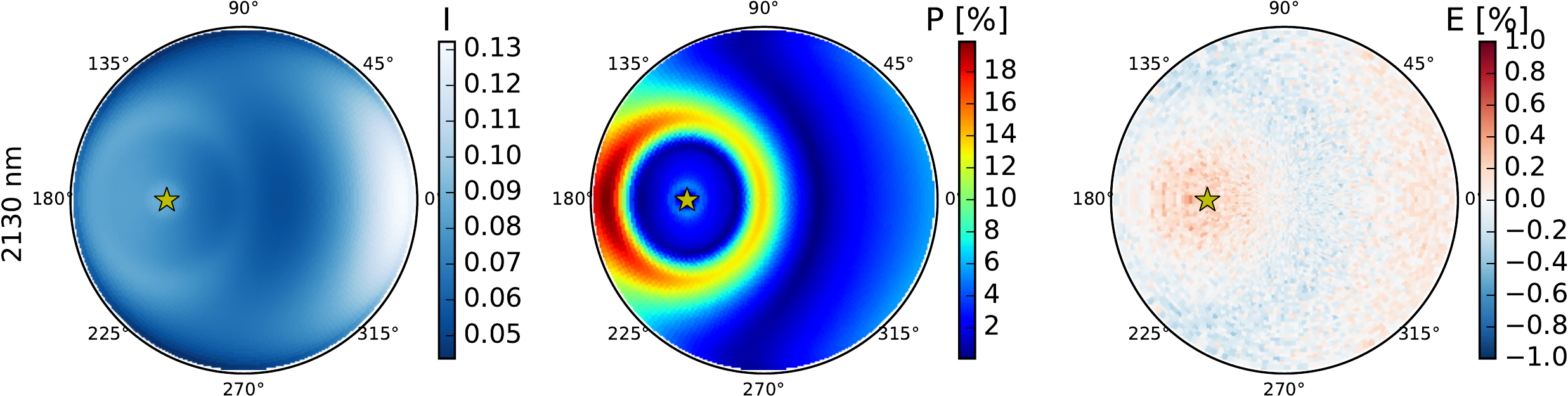}\\
  \caption{Normalized radiance, degree of polarization and error of scalar
    approximation for a down-looking observer at 120\,km altitude. The
    sun position is ($\theta_0, \phi_0$)=(40\degree, 0\degree) and the
    yellow star shows the viewing direction where the sun is in the back
    of the sensor ($\theta, \phi$)=(140\degree, 180\degree). 
    A cloud layer with an optical thickness of 6 and droplets with an
    effective radius of 10~$\mu$m has been included at an altitude
    from 2--3\,km into the US-standard molecular atmosphere. 
    The rows correspond to wavelengths of 400\,nm, 660\,nm and
    2130\,nm. 
  }  \label{fig:error_polarplot_cloud}
\end{figure*}
Figure~\ref{fig:error_polarplot_cloud} shows the radiance, the degree
of polarization and the error due to the 
neglect of polarization for a one-dimensional cloud layer. The rows
correspond to wavelengths of 400\,nm, 660\,nm, and 2130\,nm. 
The cloud consists of liquid water droplets with an effective radius of
10$\mu$m, is included in the layer from 2--3\,km into the US-standard
atmosphere, and it has an optical thickness of 6. The intensity and
the degree of polarization show the strongly polarized cloudbow.
At 400\,nm the pattern is very similar as in
Figure\,\ref{fig:error_polarplot}, thus the error is clearly dominated
by Rayleigh scattering. The
magnitude of the error is reduced from the range of $\pm$8\% to about
$\pm$3\% due to multiple scattering. 

At larger wavelengths (660\,nm and 2130\,nm)
the angular pattern of the error changes:
We find an overestimation due to the neglect of
polarization around the backscattering direction
and an underestimation in the cloudbow region.
At scattering angles larger than about 90\degree\ the
scalar approximation slightly overestimates the radiance. Overall, for
wavelengths of 660\,nm and 2130\,nm,
the error is well below 1\% for all viewing
directions. For 95\% of viewing directions
the error is $<$0.3\% at 660\,nm, and $<$0.2\% at 2130\,nm. 
This result is not consistent with the
calculations by \citet{yi2014} who investigated the impact of the
neglect of polarization in radiative transfer calculations on cloud
microphysical properties from MODIS observations:
A closer look at Figure 4 (upper panel) in
\citet{yi2014} reveals that for a cloud layer with an optical
thickness of 6 and an effective droplet radius of 10\,$\mu$m the
differences between their vector and scalar radiance calculations at
650\,nm is about 2\%, certainly larger than the differences in our 1D
cloud simulations. In order to explain these differences a radiative
transfer model intercomparison would be required. 

\begin{figure}
  \centering
  \includegraphics[width=1.0\hsize]{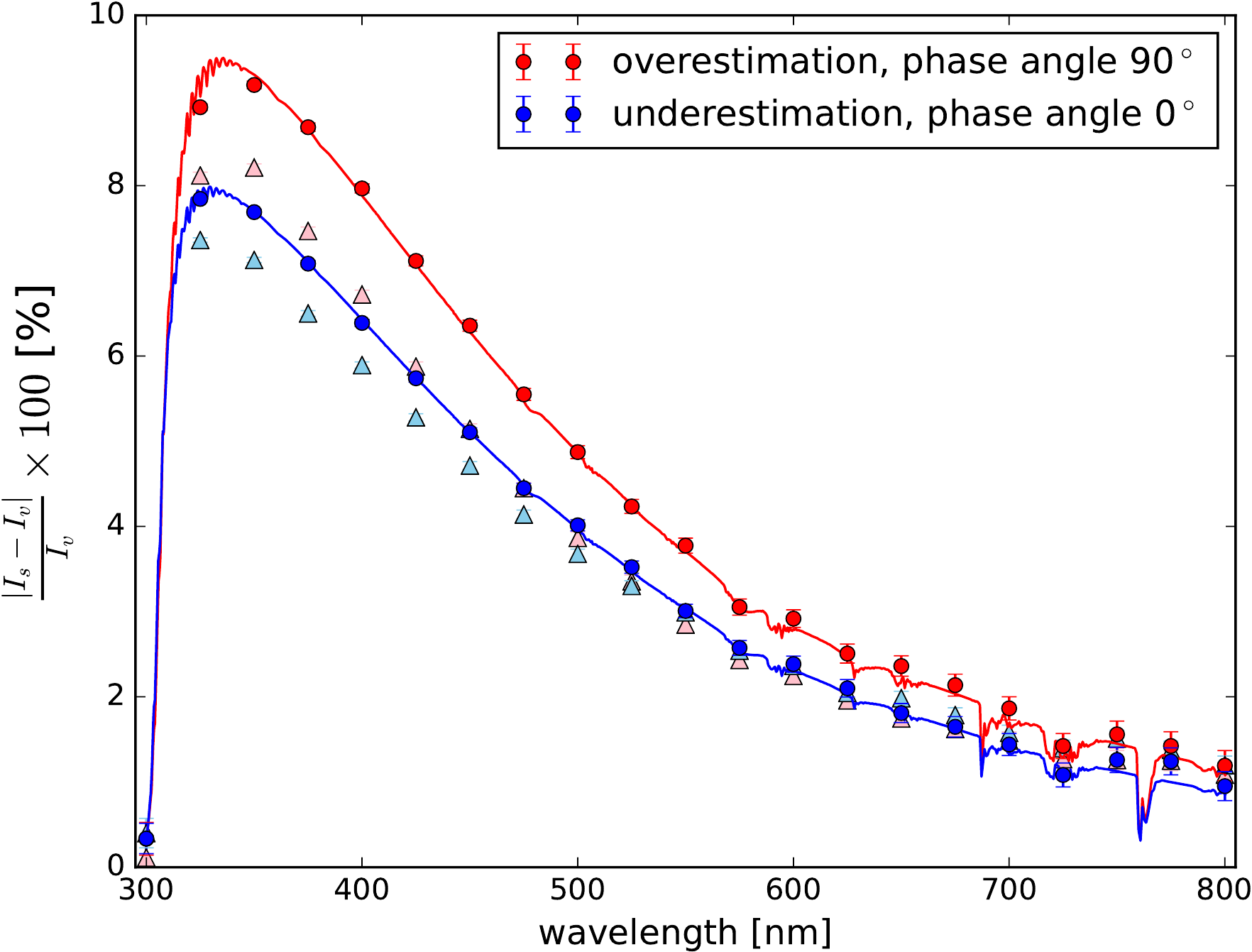}
  \caption{Maximum errors (overestimation in red and underestimation
    in blue) 
    by the scalar approximation for
    US-standard atmosphere and a solar zenith angle of 40\degree\
    as a function of wavelength. The circles
    are simulations for a pure molecular atmosphere and the triangles
    are simulations with additional aerosols (continental average from
    OPAC). The small error bars correspond to 2 standard deviations of
    the Monte Carlo results.}
  \label{fig:rayleigh_spectrum}
\end{figure}
Figure~\ref{fig:rayleigh_spectrum} shows the spectral dependence of
the maximum overestimation and the maximum underestimation due to the
scalar approximation. As in the previous simulation,
we set the solar zenith angle to 40\degree\ and the solar
azimuth angle to 180\degree. 
The observer is located at the top of the
atmosphere and we simulate two viewing directions given by the 
viewing zenith angle $\theta$ and the viewing azimuth angle $\phi$:
(1) $(\theta,\phi)$=(50\degree,
0\degree) -- for this direction the phase angle is 90\degree. 
(2) $(\theta,\phi)$=(40\degree, 180\degree) -- this direction
corresponds to backscattering (phase angle 0\degree),
i.e. the sun is behind the observer.
We choose these directions to estimate the maximum errors to be
expected for Rayleigh scattering. 
The circles represent
monochromatic simulations and the errorbars correspond to two standard
deviations of the Monte Carlo results.
In order to obtain the continuous spectrum including
absorption features (solid lines)
we have applied the absorption lines importance
sampling method \citep{emde2011}. 
We find the maximum overestimation error of 9.5\% at 335.5\,nm and the
maximum underestimation error of 8\% at 329.5\,nm. For smaller
wavelenths the errors decrease due to strong ozone absorption and due to an increasing amount of multiple scattering and for
larger wavelengths they decrease with decreasing Rayleigh scattering
coefficient. The errors are also decreased in  
the absorption bands (e.g., oxygen-A absorption about 760\,nm). 
Those results are consistent with \citet{mishchenko1994}, who
obtained decreasing errors for decreasing Rayleigh scattering optical
thicknesses and decreasing single scattering albedos.
The triangles in the figure show simulations including the US-standard
atmosphere and in addition standard continental aerosol
particles. We used the ``continental average'' aerosol as  
defined in the OPAC database and included in the
libRadtran radiative transfer package \citep{hess1998, emde2016}. We
find that the errors are only slightly reduced in presence of aerosols but are still
significant (in the range between 8\% at 350\,nm and 1.5\% at
800\,nm). 

\citet{Kotchenova2006} have shown that for
longer wavelengths and higher aerosol load, the error for aerosol scattering
can be larger than for pure Rayleigh scattering. They find
errors up to 5.3\% at 670\,nm wavelengths for an optically thick
($\tau$=0.728) biomass burning smoke aerosol model. The focus of this
study is the error for realistic 3D atmospheres including clouds, therefore
aerosols are not studied in more detail here.  

\subsection{Results for three-dimensional cloud field} 

Three-dimensional simulations are performed for 
the same sun-observer geometry as for
the spectral calculation shown in the previous section and in addition
for a phase angle of 40\degree\ where cloud scattering produces high
polarization (cloudbow, compare also Figure~\ref{fig:phase_matrix}).
In the following we show
simulations performed at a wavelength of 400~nm.

\begin{figure*}
  \centering
  \includegraphics[width=.8\hsize]{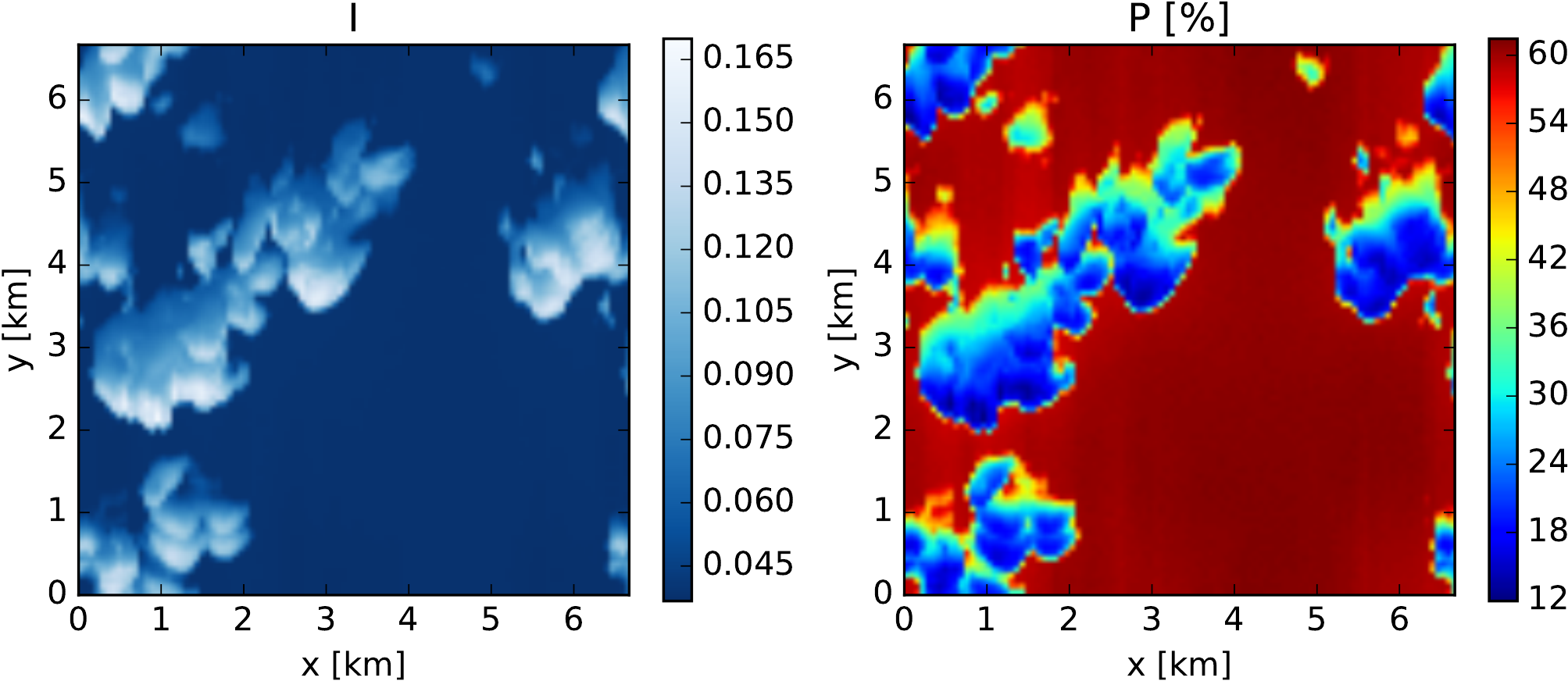}\\
  \includegraphics[width=1.0\hsize]{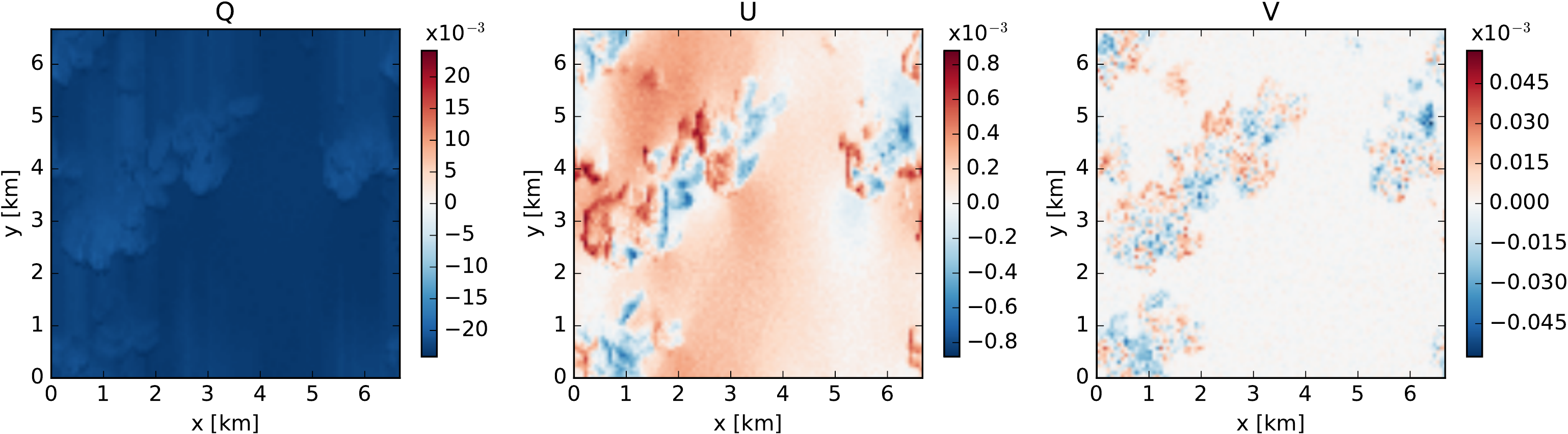}
  \caption{{\sl Top:} Normalized Radiance and degree of polarization for a down-looking
    observer at 120\,km altitude. The sun position is
    $(\theta_0,\phi_0)=(40\degree,90\degree)$ and the viewing direction
    is $(\theta,\phi)=(50\degree,270\degree)$, this corresponds to a
    phase angle of 90\degree. {\sl Bottom:} Stokes vector components
    $Q$, $U$, and $V$.} 
  \label{fig:result_rayleigh_1}
\end{figure*}
Figure~\ref{fig:result_rayleigh_1} shows the result for a phase
angle of 90\degree\, whereabouts we expect the maximum overestimation.
The sun position is $(\theta_0,\phi_0)=(40\degree,90\degree)$ and the
viewing direction is $(\theta,\phi)=(50\degree,270\degree)$.
The upper left image shows the
radiance $I$, which is increased on the illuminated sides of the
clouds. Note that the direction of incident solar radiation is along
the positive $y$-direction. 
The upper right image shows the degree of polarization, which
is in the clear-sky region about 60\%. Within the clouds it decreases
to about 15\% in illuminated regions and to about 35\% in shadowed
regions.
The bottom images show the Stokes components $Q$, $U$, and $V$. Since
the observation direction is in the solar principal plane, we find the
major amount
of polarization in the $Q$-component, which is negative, thus the
radiation is polarized perpendicular to the scattering plane which
we expect for Rayleigh scattering at 90\degree\ scattering
angle. The cloud scattering contribution to $Q$ is small in this
sun-observer geometry, thus the clouds are barely visible in the
$Q$-image. The
components $U$ and $V$ would be exactly 0 for plane-parallel
simulations for symmetry reasons. Since the 3D clouds are not
symmetric about the scattering plane we obtain polarization patterns
for $U$ and $V$. For $U$ we see that the polarization propagates from
the clouds into the clear region by multiple scattering. $U$ has
positive and negative values being 1--2 orders of magnitude smaller
than $Q$. We find small circular polarization in the clouds, however
$V$ is almost 3 orders of magnitue smaller than $Q$. 
It should be noted, that the strong decrease of the degree of
polarization in the cloud can mainly be attributed to the increase of the
radiance $I$.

\begin{figure*}
  \centering
  \includegraphics[width=1.\hsize]{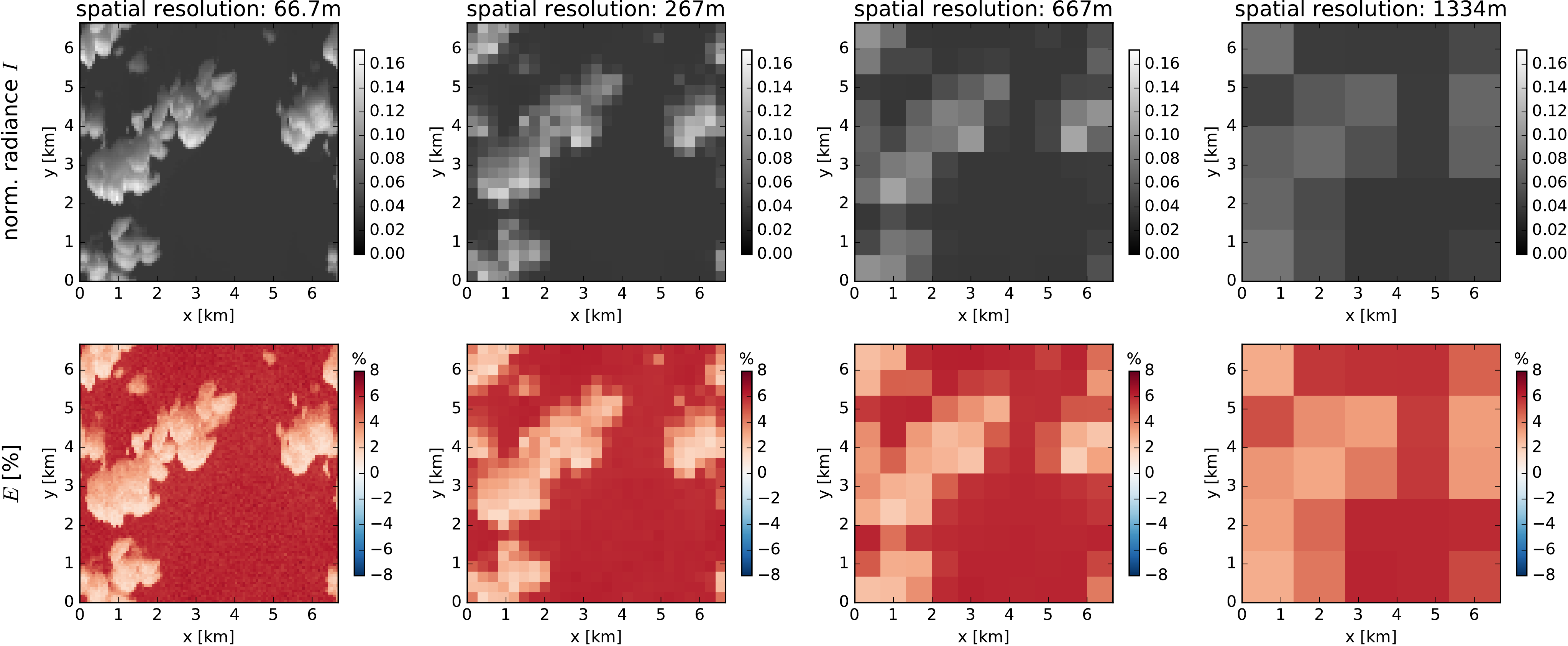}
  \caption{Normalized radiance (top) and error due to neglect of polarization (bottom) for
    various spatial resolutions. The simulations are for a phase
    angle of 90\degree, where the scalar approximation yields an
    overestimation of the radiance $I$.} 
  \label{fig:resolutions_rayleigh_1}
\end{figure*}
Figure~\ref{fig:resolutions_rayleigh_1} shows the radiance field (top
panels) and the error (overestimation) 
due to the neglect of polarization (bottom
panels). The left figure is simulated with the same spatial resolution
as the input cloud field (66.7\,m\,$\times$\,66.7\,m). The result shows,
that in the clear-sky region between the clouds, the overestimation is
about 6\%. The simulation without clouds (compare previous section)
yields an overestimation of about 8\%, thus cloud scattering into
the clearsky regions reduces the error by about 2\%. In the cloud
shadow regions the overestimation is of the order of 4\%. To assess
the question, whether it is important to consider polarization in
radiative transfer simulations for satellite remote sensing instruments,
we average the result to mimic coarser resolutions: we calculate block
mean averages of 4$\times$4, 10$\times$10 and 20$\times$20 pixels,
respectively. This way we obtain similar spatial resolutions as for
example the MODIS instrument 
with spatial resolutions of 250\,m (e.g. for band~1, bandwidth
620--670\,nm), 500~m (e.g. band 3, bandwith 459--479\,nm) and 1000~m (e.g. band 8,
bandwidth 405--420\,nm).  In the averaged radiance images the fine cumulus 
cloud structures are no longer well resolved. The error patterns are
similar to the image with highest resolution, with errors ranging from
about 4\% in cloud shadow regions to about 6\% in the clear sky
region. Many of the pixels
are partially cloudy with errors ranging from about
2--5.5\%. Thus the neglect of polarization
can cause significant errors at short wavelengths, 
in particular when the observed pixels are mostly
partially cloudy or include cloud shadows.

\begin{figure*} \centering
  \includegraphics[width=.8\hsize]{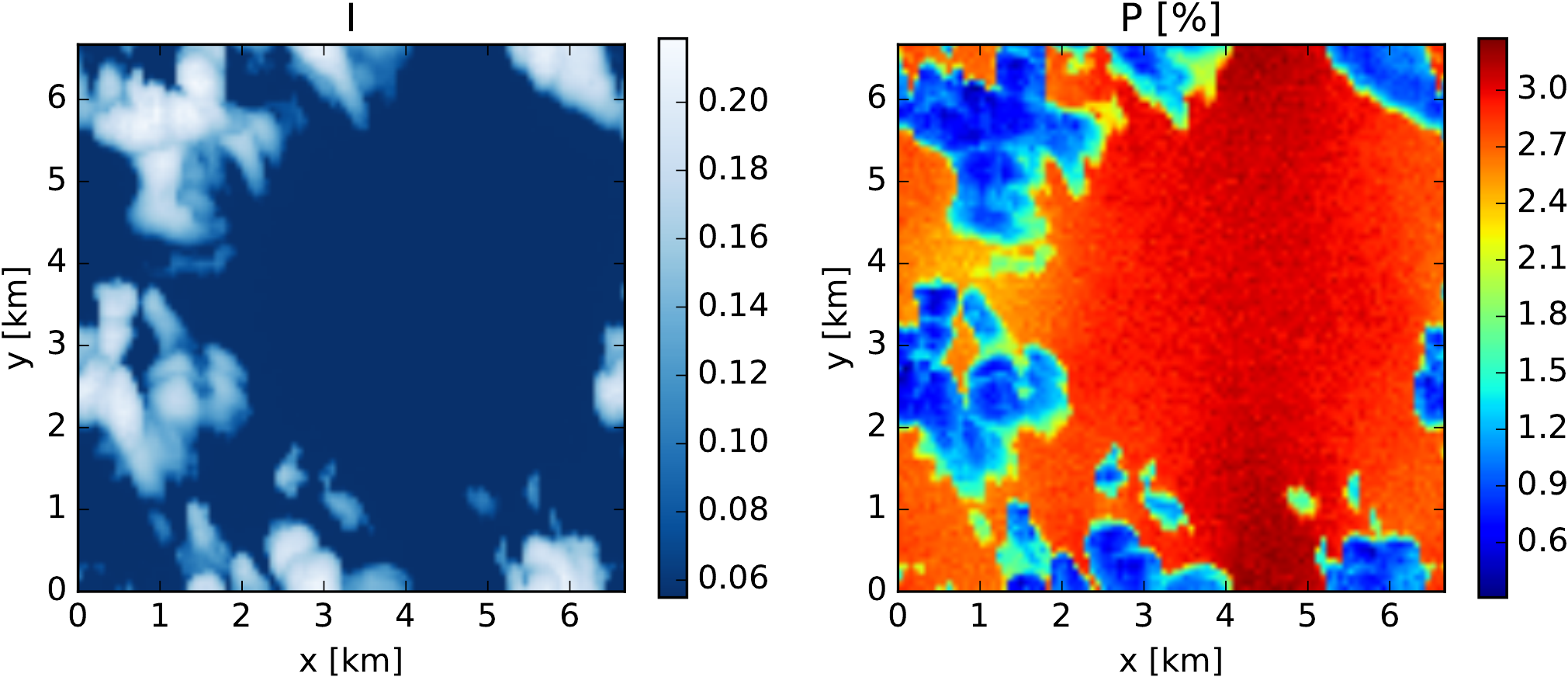}
  \includegraphics[width=1.0\hsize]{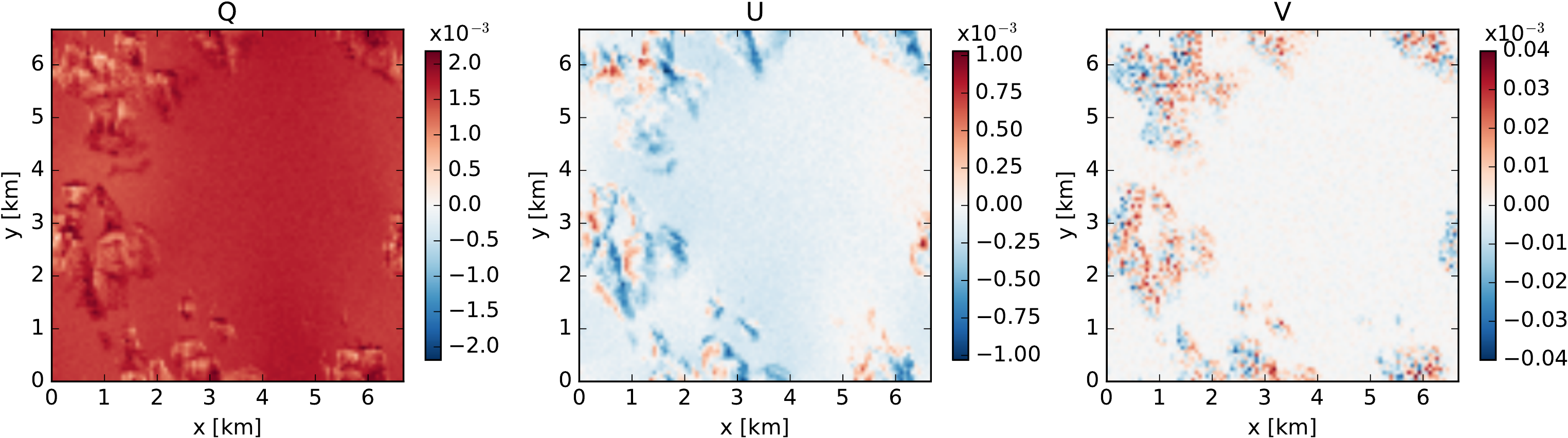}
  \caption{{\sl Top:} 
    Normalized radiance and degree of polarization for an observer at the
    top of the atmosphere
    ($(\theta_0,\phi_0)=(40\degree,90\degree),(\theta,\phi)=(40\degree,90\degree)$). The
    phase angle is 0\degree (backscattering direction). {\sl Bottom:}
    Stokes vector components $Q$, $U$, and $V$.}
  \label{fig:result_rayleigh_2}
\end{figure*}
The upper plots in 
Figure~\ref{fig:result_rayleigh_2} show the radiance field
and the degree of polarization in backscattering direction. In
the radiance image the clouds appear bright and in this geometry the
clouds do not shadow other parts of the clouds as much as 
for a phase angle of
90\degree\ (compare Figure~\ref{fig:result_rayleigh_1}).
The degree of polarization in backscattering direction is
very small, about 3\% in the clear regions and below 1\% in cloudy
regions. In regions with optically thin clouds the degree of
polarization is between 1\% and 3\% and for optically thicker clouds
it becomes close to 0.  
In backscattering direction $Q$ is positive, thus the
radiation is polarized parallel to the scattering plane as expected
for Rayleigh scattering. The clouds partly increase and partly
decrease $Q$. The Stokes component $U$, which is exactly 0 for
backscattering in a one-dimensional model, shows a pattern generated
by scattering in the three-dimensional clouds. Further, small circular
polarization (non-zero $V$) can be observed in the clouds. 

\begin{figure*} \centering
  \includegraphics[width=1.\hsize]{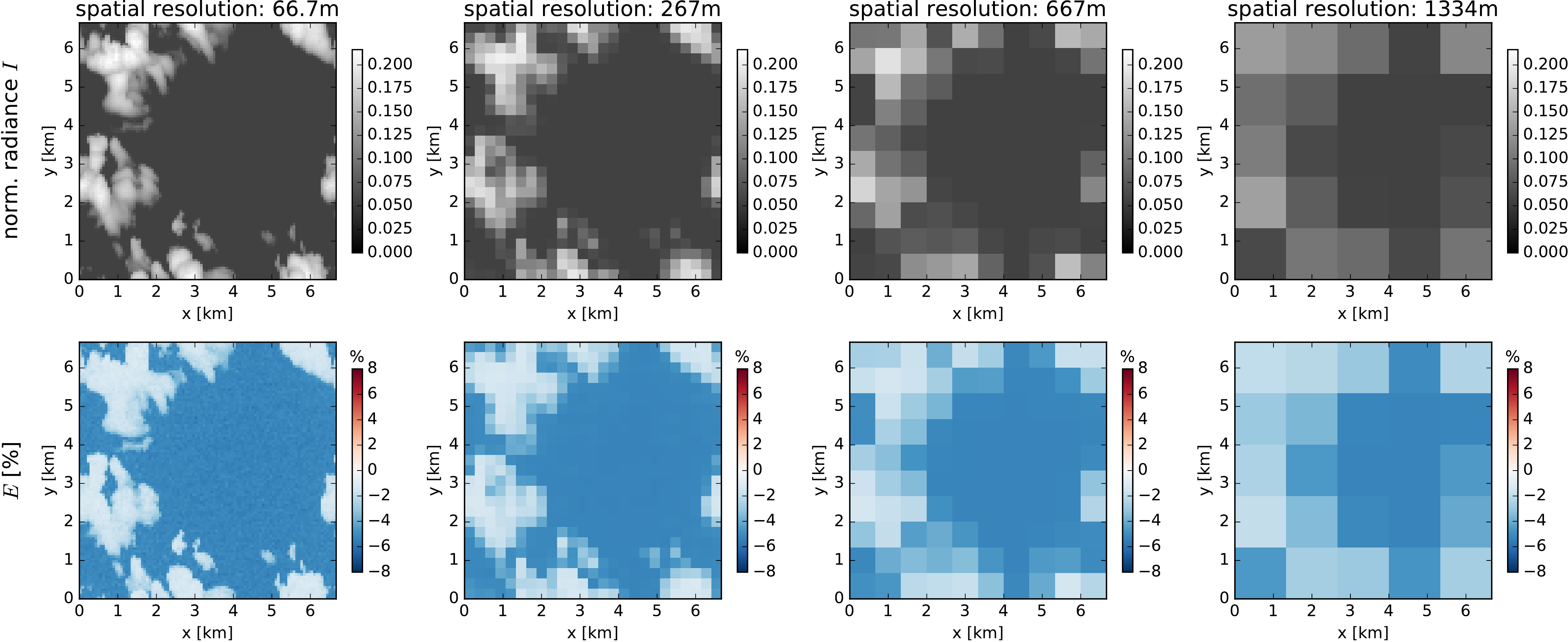}
  \caption{Normalized radiance (top) and error due to neglect of polarization (bottom) for
    various spatial resolutions. The simulations are for a phase
    angle of 0\degree, where the scalar approximation yields an
    underestimation of the radiance $I$.}
  \label{fig:resolutions_rayleigh_2}
\end{figure*}
Figure~\ref{fig:resolutions_rayleigh_2} shows the underestimation due
to the scalar approximation for various spatial resolutions
in backscattering direction. In the clear
sky regions it is about 5\%, this is about 1.5\% less compared to a
pure Rayleigh atmosphere. In the cloudy regions the error is mostly 
below 2\%. For partially cloudy pixels the error of the scalar
approximation is significant, depending on the cloud cover we
find values between 3 and 5\%. 

\begin{figure*} \centering
  \includegraphics[width=.8\hsize]{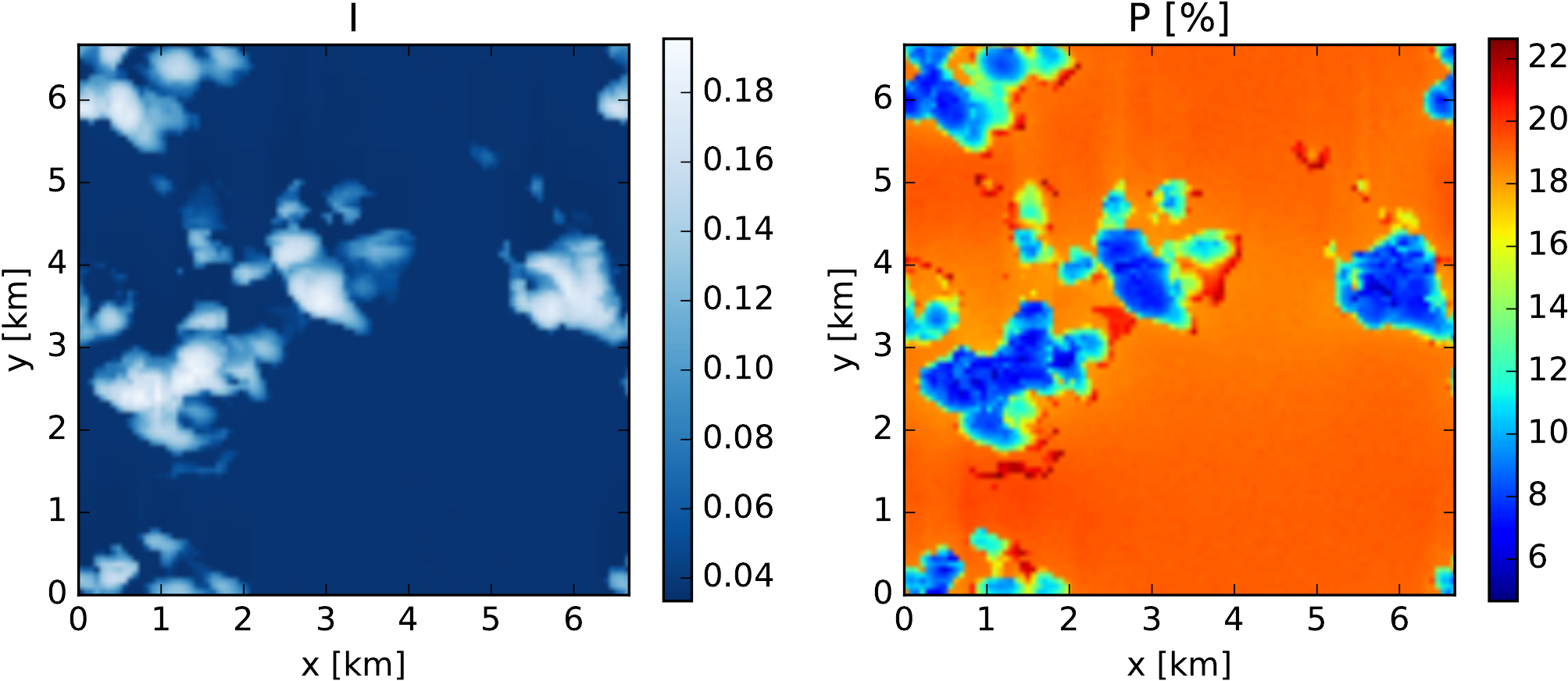}
  \includegraphics[width=1.0\hsize]{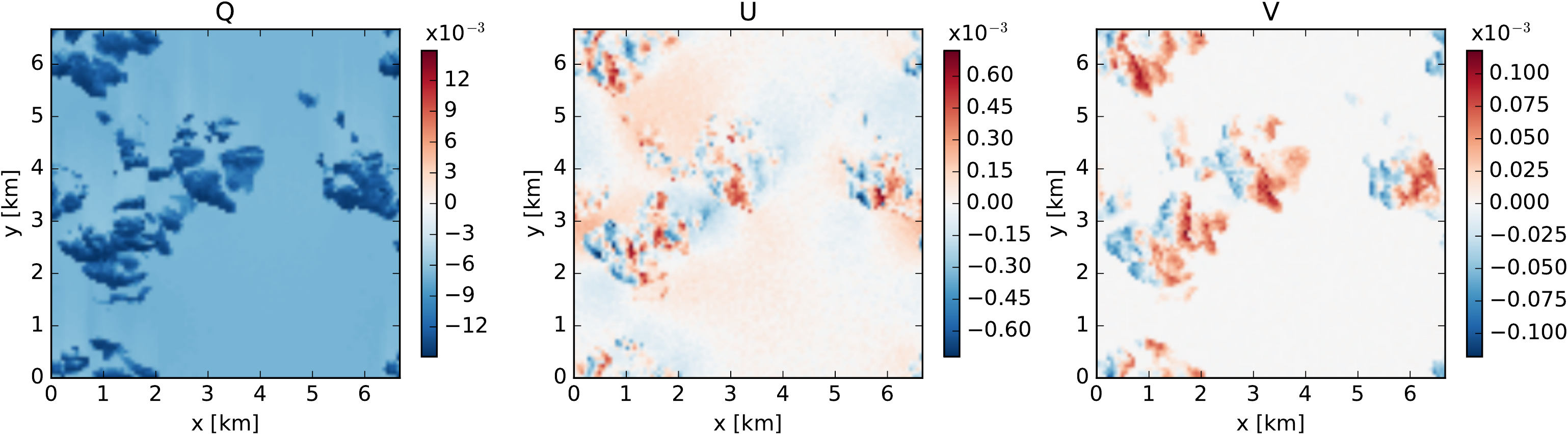}
  \caption{{\sl Top:} 
    Normalized radiance and degree of polarization for an observer at the
    top of the atmosphere
    ($(\theta_0,\phi_0)=(40\degree,90\degree),(\theta,\phi)=(0\degree,0\degree)$). The
    phase angle is 40\degree (cloudbow direction). {\sl Bottom:}
    Stokes components $Q$, $U$, and $V$. }
  \label{fig:result_rayleigh_3}
\end{figure*}
The upper images in Figure~\ref{fig:result_rayleigh_3} show the radiance field
and the degree of polarization for a phase angle of 40\degree,
corresponding to the cloudbow, where cloud scattering is highly
polarized. This can be seen in the image of $Q$ (lower left plot),
the absolute value of which is much larger in the clouds than
in the clearsky background. Nevertheless the degree of polarization is
mostly smaller in the clouds than in the clear regions, because the
increase of $I$ by multiple scattering is larger than the increase of
$Q$ which depends on fewer orders of scattering. Only for very thin
clouds the degree of polarization is increased compared to clear sky.
The viewing direction
is nadir, still we find non-zero $U$ and $V$ patterns due to the
asymmetric cloud structures.

\begin{figure*} \centering
  \includegraphics[width=1.\hsize]{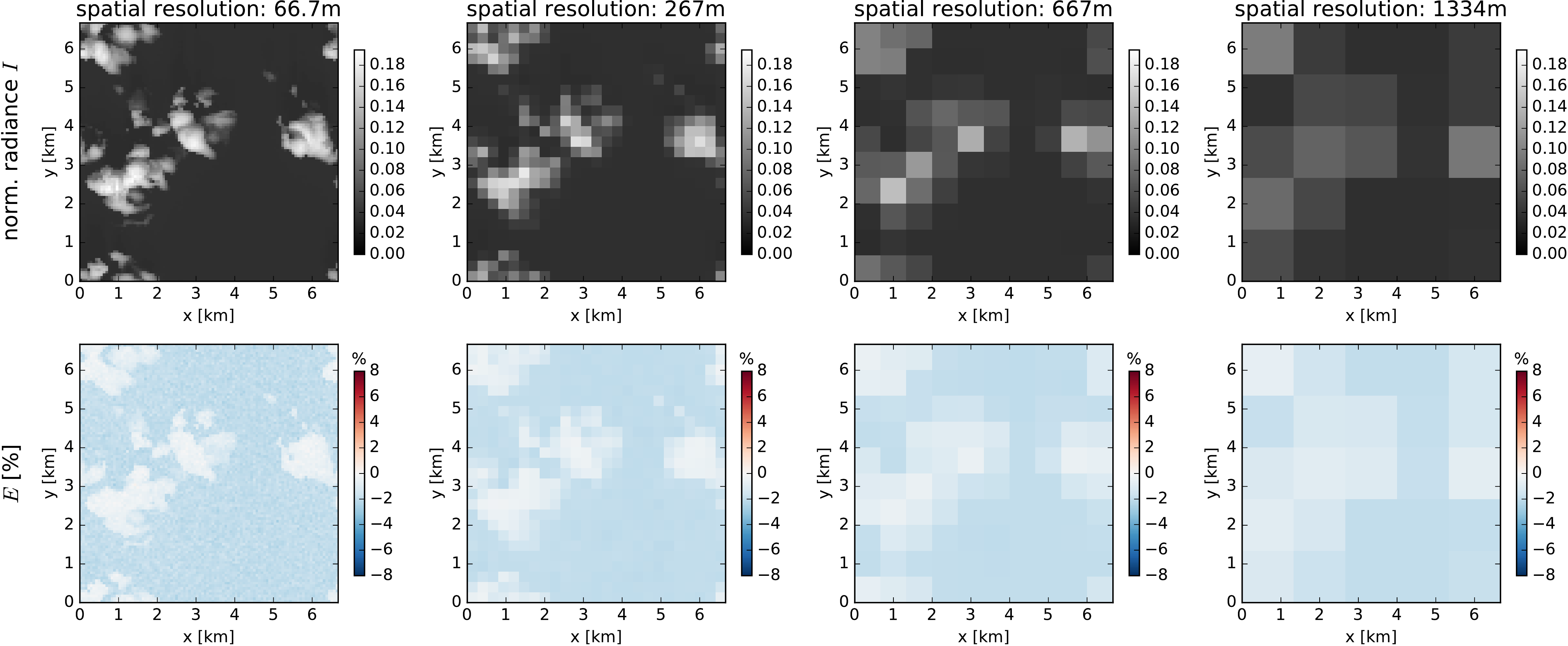}
  \caption{Normalized radiance (top) and error due to neglect of polarization (bottom) for
    various spatial resolutions. The simulations are for a phase
    angle of 40\degree, where the scalar approximation yields an
    underestimation of the radiance $I$.}
  \label{fig:resolutions_rayleigh_3}
\end{figure*}
Fig.~\ref{fig:resolutions_rayleigh_3} shows the radiance and the error
due to the neglect of polarization for various spatial
resolutions. We find that the radiance is underestimated, in the clear
region by approximately 2\%. Corresponding 1D simulations yield an
underestimation of about 2.7\%. In the cloudy and partially cloudy
parts of the image the error is always below 2\%. 
These results show, that the error due to the neglect of polarization
is caused by Rayleigh scattering, even in a sun-observer geometrie
where cloud scattering causes significant polarization. 

We have performed the simulations in 
cloudbow geometry for wavelengths relevant for
cloud remote sensing, namely 660\,nm and 2130\,nm (not shown).
For 660\,nm we
obtain a maximum underestimation of 0.7\% in the clear part of the
image and even smaller errors in the cloudy regions. 
For 2130\,nm the error
is below 0.5\% in the whole image. 

\section{Summary and conclusions}
\label{sec:conclusions}

We have investigated the error due to the neglect of
polarization in radiative transfer calculations for a realistic
three-dimensional atmosphere. 

First, in order to get an idea of the error distribution depending on
viewing direction and sensor position,
we have calculated the error for all viewing
directions from the surface as well as from space.
We have used 
the US-standard (multi-layer) atmosphere, and performed radiative
transfer simulations at a wavelength of 400\,nm
in full vector and in scalar mode, respectively,
and calculate the relative difference between the obtained radiances. 
In accordance to \citet{mishchenko1994} we find 
the maximum overestimation of about 8\% for a
phase angle of 90\degree\ and the maximum underestimation of about
6.5\% for phase angles of 0\degree\ or 180\degree.
For other viewing directions, the
error is between these two extreme values. 

The same simulations including a homogeneous cloud layer yield errors
smaller than 3\% at 400\,nm for all sun-observer geometries, including the cloudbow
region where cloud scattering causes high polarization. For larger
wavelengths (660\,nm and 2130\,nm) the errors are generally well below 1\%. 

Next, we investigated the spectral dependance of the maximum 
over-/underestimation. As expected we find that it decreases from
about 9.5\%/7.5\%  at 350\,nm to about 1.3\%/1\% at 800\,nm as the Rayleigh
optical thickness decreases. In the water vapor and oxygen absorption
bands the error is also decreased, this result is again consistent
with \citet{mishchenko1994} who showed that the error decreases with
decreasing single scattering albedo. 
We found that for typical continental 
average aerosol conditions the error is  decreased
by 1--2\% at shorter wavelengths up to about 500\,nm.
At longer wavelengths the magnitude of the 
error is not changed by this type of aerosol.

Finally, we performed simulations 
for a three-dimensional cloud field
surrounded by the US-standard molecular atmosphere.
The spatial resolution of the cloud field is 66.7\,m\,$\times$\,66.7\,m
and the domain size is 6.67\,km\,$\times$\,6.67\,km. We simulated
100$\times$100 pixels for a sensor located at the top of the
atmosphere. 

In order to obtain the maximum errors that can be expected
due to Rayleigh scattering, we simulated two
viewing directions corresponding to phase angles of 90\degree\ and
0\degree, respectively. In the clear regions between the clouds we find that the
overestimation error is about 
6\%, which is approximately 2\% less than in the pure clear-sky
atmosphere without surrounding clouds. The reason for this decrease is
that photons which are multiple scattered in the clouds enter the
clear-sky region without a preferred direction. When those photons are
scattered by molecules towards the observer, they do not have a
preferred polarization direction. 
Within the clouds the error is still up to 4\% with largest values in
shadowed regions. In order to simulate coarser spatial resolutions
including partially cloudy pixels we spatially averaged our results over
267\,m, 667\,m, and 1334\,m, respectively. 
Those results are typical for satellite observations with coarser
resolutions. We find that in the partially cloudy pixels the
overestimation error ranges from 2-5.5\%. 
The underestimation errors are generally
slightly smaller than the overestimation errors.

In order to test whether scattering at cloud droplets cause significant
errors we performed 3D simulations at a phase angle of 40\degree\
corresponding to the highly polarized cloudbow. At 400\,nm, where the
Rayleigh scattering contribution is large, we find errors of up to 2\%
in the region between the clouds and smaller errors within the
clouds. At 660\,nm and 2130\,nm, wavelengths that are typically used
for cloud remote sensing, the errors are less than 1\% in the full domain.

In summary, our results show that the error due to the neglect of
polarization is
not always negligible for radiative transfer in realistic
clouds at shorter wavelengths with significant Rayleigh scattering
contribution.  As shown in
Figure~\ref{fig:rayleigh_spectrum} the error decreases with
wavelengths as the amount of Rayleigh scattering decreases.  

Many cloud retrieval algorithms utilize longer wavelengths, as for instance
the widely used lookup-table method by \citet{nakajima1990}, which has
been developed for observations at 750\,nm and at 2160\,nm. For the
operational cloud properties retrieval algorithms from MODIS
observations this method has been adapted for various channel
combinations using spectral bands in the range from 0.67\,$\mu$m to
3.7\,$\mu$m \citep{platnick2017}.  Our results suggest that for this
type of cloud retrieval
algorithms the scalar approximation can safely be applied.
However, \citet{yi2014} found that the neglect of polarization in the
radiative transfer simulations produce errors as large as 15\% in
retrieved cloud optical thickness and effective droplet radius. They
do not analyse in detail the cases which produce the very large
errors. In our one-dimensional radiative transfer simulations we did
not find sun-observer geometries which would produce such large errors.

For hyperspectral retrieval methods the error might be relevant.
E.g., the algorithm by \citet{zinner2016} 
utilizes the spectral slope of the radiance between 485 and 560\,nm. In
this range the maximum error due to the scalar approximation for Rayleigh
scattering changes from
about 6\% to 4\%, so that the spectral slope will be affected by the scalar
approximation.

The scalar approximation could also have a significant impact on 
aerosol and cloud retrieval algorithms for multi-angle multi-spectral 
observations (e.g., the Multi-angle Imaging SpectroRadiometer MISR),
because the error strongly depends on observation direction; for some
directions the error is negligible while for others it might be several
per cent, in particular for wavelengths below 500\,nm that are used
for aerosol remote sensing.

\bibliographystyle{elsarticle-num-names}
\bibliography{./literature.bib}

\end{document}